\documentclass{article}
\usepackage{amssymb}
\begin{document}
\title{Effect of the variational symmetries of the Lagrangian on the propagator and associated conserved operators}

\author{G.F.\ Torres del Castillo\\
Departamento de F\'isica Matem\'atica, Instituto de Ciencias \\
Universidad Aut\'onoma de Puebla, 72570 Puebla, Pue., M\'exico\\[1ex]
R.L.\ Lechuga-Sol\'is\\
Facultad de Ciencias F\'isico Matem\'aticas \\ Universidad Aut\'onoma de Puebla, 72570 Puebla, Pue., M\'exico}

\maketitle

\begin{abstract}
Making use of the expression for the propagator in terms of path integrals, we study the effect of certain variational symmetries of a Lagrangian on the corresponding propagator. We also show that by considering a point transformation that relates two different Lagrangians one can obtain a relation between the corresponding propagators.
\end{abstract}



\section{Introduction}
In (classical) analytical mechanics, certain families of transformations are related with conserved quantities. In the Lagrangian formulation, each one-parameter family of variational symmetries of the Lagrangian leads to a constant of motion (see, {\it e.g.}, Ref.\ \cite{AB} and the references cited therein), while in the Hamiltonian formulation, each one-parameter family of canonical transformations that leave the Hamiltonian invariant is associated with a constant of motion. This last result has an analog in quantum mechanics, where each one-parameter family of unitary operators that leave invariant the Hamiltonian operator leads to a conserved operator \cite{He}. This analogy might be expected owing to the close relationship between the Hamiltonian formulation of classical mechanics and the standard formalism of the non-relativistic quantum mechanics.

However, the quantum dynamics can be also related directly with the Lagrangian formalism through the expression for the time evolution operator (propagator) in terms of path integrals (see, {\it e.g.}, Refs.\ \cite{FH,KL,GY,MK})

For example, in the case of a quantum system formed by a spin-0 particle in the three-dimensional space, the propagator, $K({\bf r}_{1}, t_{1}; {\bf r}_{0}, t_{0})$, which is the wavefunction at the point ${\bf r}_{1}$, at time $t_{1}$, if the particle was localized at ${\bf r}_{0}$, at time $t_{0}$, is related to the time evolution operator, $U(t_{1}, t_{0})$, by means of
\begin{equation}
K({\bf r}_{1}, t_{1}; {\bf r}_{0}, t_{0}) = \langle {\bf r}_{1} | U(t_{1}, t_{0}) | {\bf r}_{0} \rangle \label{prop}
\end{equation}
and, as is well known, Feynman's rule allows us to find the propagator through
\begin{equation}
K({\bf r}_{1}, t_{1}; {\bf r}_{0}, t_{0}) = \int_{({\bf r}_{0}, t_{0})}^{({\bf r}_{1}, t_{1})} {\mathcal D} \{ {\bf r}(t) \} \exp \frac{\rm i}{\hbar} \int_{t_{0}}^{t_{1}} L({\bf r}, \dot{\bf r}, t) {\rm d} t, \label{feyn}
\end{equation}
where $L({\bf r}, \dot{\bf r}, t)$ is the Lagrangian of the classical system (we do not need to be more specific about this expression, because in the rest of this paper we will not use it for explicit calculations). This relation suggests that the symmetries of the Lagrangian determine symmetries of the propagator and, therefore, that the variational symmetries of the Lagrangian must determine some conserved operators of the quantized system. One of the aims of this paper is to study such connections.

In quantum mechanics, the symmetries and the conserved quantities are usually defined making use of the Hamiltonian operator: A unitary operator, $T$ (which may depend on the time and on one or several parameters), is a symmetry (of the quantum system, or of the Hamiltonian) if
\begin{equation}
T^{-1} H T - {\rm i} \hbar \, T^{-1} \frac{\partial T}{\partial t} = H, \label{inv}
\end{equation}
and a Hermitian operator, $A$ (which may depend explicitly on the time, in spite of the fact that we are making use of the Schr\"odinger picture), is conserved if
\begin{equation}
{\rm i} \hbar \frac{\partial A}{\partial t} + [A, H] = 0. \label{cons}
\end{equation}
Making use of these definitions one can prove that if $A$ is conserved then the unitary operators $T_{s} \equiv \exp (- {\rm i} sA/\hbar)$, for $s \in \mathbb{R}$, form a one-parameter group of symmetries of $H$ and, conversely, if the unitary operators $T_{s}$ form a one-parameter family of symmetries of $H$ then, assuming that $T_{0}$ is the identity operator, $A \equiv {\rm i} \hbar \, \partial T_{s}/\partial s|_{s = 0}$ is conserved \cite{He}. (It may be noticed that if $T$ satisfies (\ref{inv}), then it also satisfies (\ref{cons}); the essential difference between Eqs.\ (\ref{inv}) and (\ref{cons}) is that in the first case the operator $T$ must have an inverse.)

In this paper we show that Feynman's formula (\ref{feyn}) allows us to determine the effect of certain variational symmetries of $L$ on the propagator, which, in turn, allows us to find the corresponding conserved operators. In Sec.\ 2, we characterize the symmetries and the conserved operators employing the time evolution operator. In Sec.\ 3, we relate variational symmetries of the Lagrangian with conserved operators by means of the Feynman integral. In Sec.\ 4 we show that when two different Lagrangians are related by means of an appropriate point transformation, one can readily find one propagator in terms of the other. Throughout the paper we give several examples.

\section{Symmetries and conserved operators in terms of the evolution operator}
Taking into account what is meant by a symmetry operator, $T$, of a quantum system, which can depend explicitly on the time, we could take as the definition of such an operator the condition
\begin{equation}
T(t_{1}) U(t_{1}, t_{0}) = U(t_{1}, t_{0}) T(t_{0}), \label{invev}
\end{equation}
for all values of $t_{0}$ and $t_{1}$ (roughly speaking, if $T$ is a symmetry operator, then one should get the same result by applying $T$ at some initial time and then allowing the system to evolve, or letting first the system to evolve and then applying the transformation $T$). In fact, we can readily demonstrate that Eq.\ (\ref{invev}) is equivalent to Eq.\ (\ref{inv}): Differentiating both sides of (\ref{invev}) with respect to $t_{1}$ and evaluating these derivatives at $t_{0} = t_{1}$, we obtain
\[
\left[ \frac{\partial T(t_{1})}{\partial t_{1}} U(t_{1}, t_{0}) + T(t_{1}) \frac{\partial U(t_{1}, t_{0})}{\partial t_{1}} \right]_{t_{0} = t_{1}} = \left[ \frac{\partial U(t_{1}, t_{0})}{\partial t_{1}} T(t_{0}) \right]_{t_{0} = t_{1}}.
\]
Making use of the fact that the time evolution operator must satisfy
\[
{\rm i} \hbar \frac{\partial U(t_{1}, t_{0})}{\partial t_{1}} = H(t_{1}) U(t_{1}, t_{0}),
\]
and that $U(t_{1}, t_{1})$ is the identity operator, we obtain (\ref{inv}).

In a similar manner one verifies that Eq.\ (\ref{cons}) is equivalent to the condition
\begin{equation}
A(t_{1}) U(t_{1}, t_{0}) = U(t_{1}, t_{0}) A(t_{0}), \label{consev}
\end{equation}
for all values of $t_{0}$ and $t_{1}$. With the aid of Eqs.\ (\ref{invev}) and (\ref{consev}), the relationship between conserved operators and symmetries of a quantum system, mentioned in the Introduction, is clearly visible.

Since the propagator is given by the matrix elements of the time evolution operator in the basis formed by the eigenstates of the position operator, Eqs.\ (\ref{invev}) and (\ref{consev}) can be written in terms of the propagator which, in turn, is related to the Lagrangian of the corresponding classical system by means of the path integral (\ref{feyn}).

\section{Effect of the variational symmetries of the Lagrangian on the propagator}
In this section we begin by recalling some basic facts about the variational symmetries of a Lagrangian. As we shall show, certain one-parameter families of variational symmetries of $L$ correspond to symmetries of the propagator.

By definition, a variational symmetry of a Lagrangian, $L(q_{i}, \dot{q}_{i}, t)$, is a coordinate transformation $q'_{i} = q'_{i}(q_{j}, t)$, $t' = t'(q_{j}, t)$, such that
\begin{equation}
L(q_{i}', \dot{q}'_{i}, t') \, \frac{{\rm d} t'}{{\rm d} t} = L(q_{i}, \dot{q}_{i}, t) + \frac{{\rm d}}{{\rm d} t} F(q_{i}, t), \label{symvar}
\end{equation}
where $\dot{q}'_{i} \equiv {\rm d} q'_{i}/{\rm d t'}$ and $F(q_{i}, t)$ is some real-valued function of $q_{i}$ and $t$, only. There exists a constant of motion associated with each one-parameter family of variational symmetries of $L$, $q'_{i} = q'_{i}(q_{j}, t, s)$, $t' = t'(q_{j}, t, s)$, where $s$ is a real parameter that takes values in some neighborhood of zero, given by
\begin{equation}
\sum_{i = 1}^{n} \frac{\partial L}{\partial \dot{q}_{i}} \eta_{i} + \xi \left( L - \sum_{i = 1}^{n} \frac{\partial L}{\partial \dot{q_{i}}} \dot{q}_{i} \right) - G, \label{com}
\end{equation}
where
\begin{equation}
\eta_{i}(q_{j}, t) \equiv \left. \frac{\partial q'_{i}}{\partial s} \right|_{s = 0}, \qquad \xi(q_{j}, t) \equiv \left. \frac{\partial t'}{\partial s} \right|_{s = 0}, \label{infgen}
\end{equation}
are the components of the so-called infinitesimal generator of the symmetry transformations, and $G(q_{i},t) \equiv \partial F/\partial s|_{s = 0}$. (We are assuming that for $s = 0$, $q'_{i}(q_{j}, t, s)$ and $t'(q_{j}, t, s)$ reduce to $q_{i}$ and $t$, respectively.) For a given Lagrangian, the functions $\eta_{i}, \xi$, and $G$, are determined by
\begin{equation}
\sum_{i = 1}^{n} \left[ \frac{\partial L}{\partial q_{i}} \eta_{i} + \frac{\partial L}{\partial \dot{q}_{i}} \left( \frac{{\rm d} \eta_{i}}{{\rm d} t} - \dot{q}_{i} \frac{{\rm d} \xi}{{\rm d} t} \right) \right] + \frac{\partial L}{\partial t} \xi + L \, \frac{{\rm d} \xi}{{\rm d} t} = \frac{{\rm d} G}{{\rm d} t} \label{invinfs}
\end{equation}
(see, {\it e.g.}, Ref.\ \cite{AB} and the references cited therein).

Taking into account that the ``path differential measure,'' ${\mathcal D} \{ {\bf r}(t) \}$, comes from integration on the coordinates and a factor that depends on the time interval, we restrict ourselves to coordinate transformations with Jacobian equal to 1 and ${\rm d} t'/{\rm d} t = 1$, so that the measure is invariant. Then, by combining Eqs.\ (\ref{feyn}) and (\ref{symvar}) we find that, under a variational symmetry of the Lagrangian satisfying these conditions, the propagator transforms according to
\[
K({\bf r}'_{1}, t'_{1}; {\bf r}'_{0}, t'_{0}) = K({\bf r}_{1}, t_{1}; {\bf r}_{0}, t_{0}) \, {\rm e}^{{\rm i}[F({\bf r}_{1}, t_{1}) - F({\bf r}_{0}, t_{0})]/\hbar}.
\]
Equivalently, in terms of the evolution operator,
\begin{equation}
\langle {\bf r}'_{1} | U(t'_{1}, t'_{0}) | {\bf r}'_{0} \rangle = \langle {\bf r}_{1} | {\rm e}^{{\rm i} F({\bf r}, t_{1})/\hbar} U(t_{1}, t_{0}) {\rm e}^{- {\rm i} F({\bf r}, t_{0})/\hbar} | {\bf r}_{0} \rangle, \label{fundam}
\end{equation}
where ${\bf r}$ is the position {\em operator}. Expressing $| {\bf r}'_{0} \rangle$ and $| {\bf r}'_{1} \rangle$ in terms of $| {\bf r}_{0} \rangle$ and $| {\bf r}_{1} \rangle$, respectively, one obtains a relation of the form (\ref{invev}), which allows us to identify a symmetry operator associated with the variational symmetry of $L$. (Note that this is true for discrete or continuous transformations.)

\subsection{Examples}
As a first simple example, we consider a particle of mass $m$ in a uniform gravitational field, in one dimension. The standard Lagrangian is
\begin{equation}
L(x, \dot{x}, t) = \frac{m}{2} \dot{x}^{2} - mgx \label{unif}
\end{equation}
(in the case of a uniform electric field, $E$, we replace $mg$ by $-eE$, where $e$ is the electric charge of the particle). This Lagrangian possesses a five-dimensional group of variational symmetries \cite{AB}. One subgroup of these variational symmetries are the Galilean transformations
\begin{equation}
x' = x - Vt, \qquad t' = t, \label{gal}
\end{equation}
where $V$ is a real parameter that takes the place of the parameter $s$ employed above. In fact, a straightforward computation shows that
\begin{eqnarray*}
L(x', \dot{x}', t') & = & \frac{m}{2} (\dot{x} - V)^{2} - mg(x - Vt) \\
& = & L(x, \dot{x}, t) + \frac{\rm d}{{\rm d} t} \left( - m V x + \frac{mV^{2}t}{2} + \frac{mgVt^{2}}{2} \right),
\end{eqnarray*}
which allows us to identify the function $F(x, t)$ appearing in Eq.\ (\ref{symvar}). Thus, in this case, Eq.\ (\ref{fundam}) takes the form
\begin{equation}
\langle x_{1} - Vt_{1} | U(t_{1}, t_{0}) | x_{0} - Vt_{0} \rangle = \langle x_{1} | {\rm e}^{{\rm i} F(x, t_{1})/\hbar} U(t_{1}, t_{0}) {\rm e}^{- {\rm i} F(x, t_{0})/\hbar} |x_{0} \rangle, \label{15}
\end{equation}
with $F(x, t) = - m V x + \frac{1}{2} mV^{2}t + \frac{1}{2} mgVt^{2}$.

Using the well-known fact that, for any real number, $a$,
\begin{equation}
|x_{0} + a \rangle = {\rm e}^{- {\rm i} a p/\hbar} |x_{0} \rangle, \label{trans}
\end{equation}
where $p$ is the momentum operator, Eq.\ (\ref{15}) can also be written as
\[
\langle x_{1}| {\rm e}^{- {\rm i} Vt_{1} p/\hbar} U(t_{1}, t_{0}) {\rm e}^{{\rm i} Vt_{0} p/\hbar} |x_{0} \rangle = \langle x_{1}| {\rm e}^{{\rm i} F(x, t_{1})/\hbar} U(t_{1}, t_{0}) {\rm e}^{- {\rm i} F(x, t_{0})/\hbar} |x_{0} \rangle,
\]
where all the reference to $x_{1}$ and $x_{0}$ (which are two arbitrary real numbers) now appears in the bra $\langle x_{1}|$ and the ket $|x_{0} \rangle$, respectively. Hence, we have the relation between {\em operators}
\[
{\rm e}^{- {\rm i} Vt_{1} p/\hbar} U(t_{1}, t_{0}) {\rm e}^{{\rm i} Vt_{0} p/\hbar} = {\rm e}^{{\rm i} F(x, t_{1})/\hbar} U(t_{1}, t_{0}) {\rm e}^{- {\rm i} F(x, t_{0})/\hbar},
\]
which amounts to
\[
{\rm e}^{- {\rm i} F(x, t_{1})/\hbar} {\rm e}^{- {\rm i} Vt_{1} p/\hbar} U(t_{1}, t_{0}) = U(t_{1}, t_{0}) {\rm e}^{- {\rm i} F(x, t_{0})/\hbar} {\rm e}^{- {\rm i} Vt_{0} p/\hbar}.
\]
Comparing this last equation with Eq.\ (\ref{invev}) we conclude that the operator
\begin{equation}
T_{V}(t) \equiv \exp \left[ \frac{\rm i}{\hbar} \left( m V x - \frac{mV^{2}t}{2} - \frac{mgVt^{2}}{2} \right) \right] \exp \left( - \frac{\rm i}{\hbar} Vt p \right) \label{trgal}
\end{equation}
is a symmetry of the system and, therefore, its infinitesimal generator
\[
A(t) \equiv {\rm i} \hbar \left. \frac{\partial T_{V}}{\partial V} \right|_{V = 0} = - mx + \frac{mgt^{2}}{2} + tp,
\]
is conserved ({\it cf.}\ Ref.\ \cite{He}). Furthermore, since $T_{V} = \exp (- {\rm i} VA/\hbar)$, it follows that (\ref{trgal}) must be equivalent to
\[
T_{V}(t) = \exp \left[ - \frac{\rm i}{\hbar} \left( - mVx + \frac{mgVt^{2}}{2} + Vtp \right) \right].
\]
Note that we did not have to know the explicit expression of the propagator or of the evolution operator.

A second example, more involved that the previous one, is given by the standard Lagrangian for a particle of mass $m$ in a uniform gravitational field,
\[
L = \frac{m}{2} (\dot{x}^{2} + \dot{y}^{2}) - mgy.
\]
As shown in Ref.\ \cite{AB}, solving Eq.\ (\ref{invinfs}) one finds that a variational symmetry of this Lagrangian is given by the functions
\begin{equation}
\xi = 0, \qquad \eta_{1} = y + {\textstyle \frac{1}{2}} g t^{2}, \qquad \eta_{2} = - x. \label{comp}
\end{equation}
The corresponding group of coordinate transformations is found to be
\begin{eqnarray}
x' & = & x \cos s + y \sin s + {\textstyle \frac{1}{2}} gt^{2} \sin s, \nonumber \\
y' & = & - x \sin s + y \cos s + {\textstyle \frac{1}{2}} gt^{2} (\cos s - 1), \label{t2} \\
t' & = & t. \nonumber
\end{eqnarray}
In fact, a straightforward computation shows that
\[
L(x', y',\dot{x}', \dot{y}', t') = L(x, y, \dot{x}, \dot{y}, t) + \frac{\rm d}{{\rm d} t} mg \left[ ty (1 - \cos s) + tx \sin s + {\textstyle \frac{1}{2}} gt^{3} (1 - \cos s) \right],
\]
which is of the form (\ref{symvar}), with
\[
F = mg \left[ ty (1 - \cos s) + tx \sin s + {\textstyle \frac{1}{2}} gt^{3} (1 - \cos s) \right].
\]

The effect of the transformation (\ref{t2}) on the eigenstates of the position operators is [{\it cf.}\ Eqs.\ (\ref{comp})]
\begin{eqnarray*}
|{\bf r}' \rangle & = & \exp \big\{ \! - {\rm i} s [(y + {\textstyle \frac{1}{2}} g t^{2}) p_{x} - x p_{y}]/\hbar \big\} \, |{\bf r} \rangle \\
& = & \exp \big[ {\rm i} s (L_{z} - {\textstyle \frac{1}{2}} g t^{2} p_{x})/\hbar \big] \, |{\bf r} \rangle,
\end{eqnarray*}
where ${\bf r} = (x, y)$ and $L_{z}$ is the angular momentum operator $x p_{y} - y p_{x}$. Hence, from Eq.\ (\ref{fundam}), we have
\[
{\rm e}^{- {\rm i} s (L_{z} - {\textstyle \frac{1}{2}} g t_{1}{}^{2} p_{x})/\hbar} U(t_{1}, t_{0}) {\rm e}^{{\rm i} s (L_{z} - {\textstyle \frac{1}{2}} g t_{0}{}^{2} p_{x})/\hbar} = {\rm e}^{{\rm i} F(x, y, t_{1})/\hbar} U(t_{1}, t_{0}) {\rm e}^{- {\rm i} F(x, y, t_{0})/\hbar},
\]
which implies that the operator
\[
T_{s} = {\rm e}^{- {\rm i} F(x, y, t)/\hbar} \, {\rm e}^{- {\rm i} s (L_{z} - {\textstyle \frac{1}{2}} g t^{2} p_{x})/\hbar}
\]
is symmetry of the system [see Eq.\ (\ref{invev})] and that
\[
A = {\rm i} \hbar \left. \frac{\partial T_{s}}{\partial s} \right|_{s = 0} = L_{z} - {\textstyle \frac{1}{2}} g t^{2} p_{x} + mgtx
\]
is conserved. (It may be noticed that, for $g = 0$, the transformations (\ref{t2}) become rotations in the $xy$-plane about the origin and, in that case, the conserved operator, $A$, is just the angular momentum, as one would expect.)

\section{Point transformations that relate two different Lagrangians}
Apart from the coordinate transformations that leave invariant a given Lagrangian, the coordinate transformations that relate two different Lagrangians are also very useful. This procedure has been applied previously to some specific examples: In Ref.\ \cite{BH} the effect of extended Galilean transformations on the one-dimensional Schr\"odinger equation is studied making use of path integrals and in Ref.\ \cite{So}, considering also Galilean transformations, the propagator for a charged particle in a crossed uniform electromagnetic field is obtained from that of a charged particle in a uniform magnetic field.

We shall consider coordinate transformations $q'_{i} = q'_{i}(q_{j}, t)$, $t' = t'(q_{j}, t)$, such that
\begin{equation}
L^{(1)}(q_{i}', \dot{q}'_{i}, t') \, \frac{{\rm d} t'}{{\rm d} t} = L^{(2)}(q_{i}, \dot{q}_{i}, t) + \frac{{\rm d}}{{\rm d} t} F(q_{i}, t), \label{nonsym}
\end{equation}
where $L^{(1)}$ and $L^{(2)}$ are two Lagrangians with the same number of degrees of freedom and $F(q_{i}, t)$ is a function of $q_{i}$ and $t$ only [{\it cf}.\ Eq.\ (\ref{symvar})]. For instance, the coordinate transformation
\begin{equation}
x' = x + {\textstyle \frac{1}{2}} gt^{2}, \qquad t' = t, \label{accel}
\end{equation}
relates the Lagrangians
\[
L^{(1)}(x, \dot{x}, t) = {\textstyle \frac{1}{2}} m \dot{x}^{2}, \qquad L^{(2)}(x, \dot{x}, t) = {\textstyle \frac{1}{2}} m \dot{x}^{2} - mgx,
\]
corresponding to a free particle and a particle in a uniform gravitational field, respectively. Indeed,
\begin{eqnarray*}
L^{(1)}(x', \dot{x}' , t') \, \frac{{\rm d} t'}{{\rm d} t} & = & {\textstyle \frac{1}{2}} m (\dot{x} + gt)^{2} \\
& = & {\textstyle \frac{1}{2}} m \dot{x}^{2} + mgt \dot{x} + {\textstyle \frac{1}{2}} mg^{2}t^{2} \\
& = & {\textstyle \frac{1}{2}} m \dot{x}^{2} - mgx + \frac{\rm d}{{\rm d} t} (mgtx + {\textstyle \frac{1}{6}} mg^{2}t^{3}),
\end{eqnarray*}
which is of the form (\ref{nonsym}), with $F(x, t) = mgtx + {\textstyle \frac{1}{6}} mg^{2}t^{3}$.

From Eq.\ (\ref{feyn}) we find that, under a coordinate transformation with Jacobian equal to 1 and ${\rm d} t'/{\rm d} t = 1$, such that Eq.\ (\ref{nonsym}) holds, the propagators are related by
\begin{equation}
K^{(1)}({\bf r}'_{1}, t'_{1}; {\bf r}'_{0}, t'_{0}) = K^{(2)}({\bf r}_{1}, t_{1}; {\bf r}_{0}, t_{0}) \exp \frac{\rm i}{\hbar} [F({\bf r}_{1}, t_{1}) - F({\bf r}_{0}, t_{0})], \label{relprop}
\end{equation}
or, in terms of the evolution operators,
\begin{equation}
\langle {\bf r}'_{1} | U^{(1)}(t'_{1}, t'_{0}) | {\bf r}'_{0} \rangle = \langle {\bf r}_{1} | {\rm e}^{{\rm i} F({\bf r}, t_{1})/\hbar} U^{(2)}(t_{1}, t_{0}) {\rm e}^{- {\rm i} F({\bf r}, t_{0})/\hbar} | {\bf r}_{0} \rangle. \label{relevo}
\end{equation}

\subsection{Examples}
As is well known, the propagator for a free particle in one dimension is
\[
K^{(1)}(x_{1}, t_{1}; x_{0}, t_{0}) = \sqrt{\frac{m}{2 \pi {\rm i} \hbar (t_{1} - t_{0})}} \exp \frac{{\rm i} m (x_{1} - x_{0})^{2}}{2 \hbar (t_{1} - t_{0})}.
\]
Hence, making use of Eqs.\ (\ref{relprop}) and (\ref{accel}) we find that the propagator for a particle in a uniform gravitational field must be given by
\begin{eqnarray*}
K^{(2)}(x_{1}, t_{1}; x_{0}, t_{0}) & = & \sqrt{\frac{m}{2 \pi {\rm i} \hbar (t_{1} - t_{0})}} \exp \frac{{\rm i} m (x'_{1} - x'_{0})^{2}}{2 \hbar (t_{1} - t_{0})} \\
& & \mbox{} \times \exp \frac{\rm i}{\hbar} \left( mgt_{0}x_{0} + {\textstyle \frac{1}{6}} mg^{2}t_{0}{}^{3} - mgt_{1}x_{1} - {\textstyle \frac{1}{6}} mg^{2}t_{1}{}^{3} \right) \\
& = & \sqrt{\frac{m}{2 \pi {\rm i} \hbar \Delta t}} \exp \frac{{\rm i} m}{2 \hbar \Delta t} \big[ (x_{1} - x_{0})^{2} - g (x_{1} + x_{0}) (\Delta t)^{2} - {\textstyle \frac{1}{12}} g^{2} (\Delta t)^{4} \big],
\end{eqnarray*}
with $\Delta t \equiv t_{1} - t_{0}$.

On the other hand, making use of Eqs.\ (\ref{trans}) and (\ref{accel}) in (\ref{relevo}) we obtain the relation between the evolution operators
\[
{\rm e}^{{\rm i} gt_{1}{}^{2} p/2 \hbar} U^{(1)}(t_{1}, t_{0}) {\rm e}^{- {\rm i} gt_{0}{}^{2} p/2 \hbar} = {\rm e}^{{\rm i} F(x, t_{1})/\hbar} U^{(2)}(t_{1}, t_{0}) {\rm e}^{- {\rm i} F(x, t_{0})/\hbar},
\]
which means that the operator
\[
{\rm e}^{- {\rm i} F(x, t)/\hbar} \, {\rm e}^{{\rm i} gt^{2} p/2 \hbar} = {\rm e}^{- {\rm i} (mgtx + {\textstyle \frac{1}{6}} mg^{2}t^{3})/\hbar} \, {\rm e}^{{\rm i} gt^{2} p/2 \hbar}
\]
maps solutions of the Schr\"odinger equation for a free particle into solutions of the Schr\"odinger equation for a particle in a uniform gravitational field ({\it cf.}\ Refs.\ \cite{BH,BC} and the references cited therein).

Another well-known example corresponds to the Lagrangians
\[
L^{(1)}(x, y, \dot{x}, \dot{y}, t) = \frac{m}{2} (\dot{x}^{2} + \dot{y}^{2}) - \frac{m \omega^{2}}{2} (x^{2} + y^{2}),
\]
of a two-dimensional isotropic harmonic oscillator, and
\[
L^{(2)}(x, y, \dot{x}, \dot{y}, t) = \frac{m}{2} (\dot{x}^{2} + \dot{y}^{2}) + \frac{eB_{0}}{2c} (x \dot{y} - y \dot{x}),
\]
of a charged particle in a uniform magnetic field perpendicular to the $xy$-plane, provided that $B_{0} = 2m \omega c/e$ (or, equivalently, $\omega = eB_{0}/2mc$). One readily verifies that the relation (\ref{nonsym}) is satisfied with
\[
x' = x \cos \omega t - y \sin \omega t, \quad y' = x \sin \omega t + y \cos \omega t,
\]
$t' = t$, and $F = 0$. Thus, we conclude that the operator $T = \exp (- {\rm i} \omega t L_{z})$ maps solutions of the Schr\"odinger equation for the two-dimensional harmonic oscillator into solutions of the Schr\"odinger equation for a particle in a uniform magnetic field.

\section{Concluding remarks}
Apart from the expression (\ref{feyn}), the propagator can also be written in terms of path integrals in the phase space (see, {\it e.g.}, Refs.\ \cite{CG,KL}) and therefore one could also consider the effect of canonical transformations on the propagator. The possibility of using canonical coordinates is highly interesting, because one can relate (locally at least) any pair of systems with the same number of degrees of freedom by means of a canonical transformation.

\end{document}